\documentclass[sigconf,10pt]{acmart}
\AtBeginDocument{%
  }

\usepackage{multicol}
\usepackage{multirow}
\usepackage{appendix}
\usepackage{caption}
\usepackage{subcaption}
\usepackage[graphicx]{realboxes}
\usepackage{booktabs}
\usepackage{url, pifont}
\usepackage{makecell}
\usepackage{amsmath,bbm}
\usepackage{appendix}
\newcommand{\etal}{\textit{et al}.}

\newcommand{\eg}{\textit{e}.\textit{g}.}

\copyrightyear{2025}
\acmYear{2025}
\setcopyright{rightsretained}
\acmConference[ACM MOBICOM '25]{The 31st Annual International Conference on Mobile Computing and Networking}{November 4--8, 2025}{Hong Kong, China}
\acmBooktitle{The 31st Annual International Conference on Mobile Computing and Networking (ACM MOBICOM '25), November 4--8, 2025, Hong Kong, China}\acmDOI{10.1145/3680207.3765659}
\acmISBN{979-8-4007-1129-9/2025/11}
\begin{document}

\title{Efficient Geometry Compression and Communication for 3D Gaussian Splatting Point Clouds}

\author{Liang Xie}
\email{{LXie5201@outlook.com}}
\affiliation{%
\institution{School of Computer Science and Technology, Guangdong University of
Technology}
\city{Guangzhou}
\country{China}
}
\affiliation{%
\institution{Shenzhen Graduate School, Peking University}
\city{Shenzhen}
\country{China}
}

\author{Yanting Li}
\email{1209847234@qq.com}
\affiliation{
  \institution{School of Computer Science and Technology, Guangdong University of Technology}
\city{Guangzhou}
\country{China}
}

\author{Luyang Tang}
\email{tly926@stu.pku.edu.cn}
\affiliation{
  \institution{Shenzhen Graduate School, Peking University}
  \city{Shenzhen}
\country{China}
}

\author{Wei Gao}
% \authornote{Corresponding author: Wei Gao}
\authornote{Corresponding author: Wei Gao. This work was supported by The Major Key Project of PCL (PCL2024A02), Natural Science Foundation of China (6227101362031013), Guangdong Provincial Key Laboratory of Ultra High Defnition Immersive Media Technology (2024B1212010006), Guangdong Province Pearl River Talent Program (2021QN020708), Guangdong Basic and Applied Basic Research Foundation (2024A1515010155), Shenzhen Science and TechnologyProgram (JCYJ20240813160202004, JCYJ20230807120808017).}
\email{gaowei262@pku.edu.cn}
\affiliation{%
 \institution{Guangdong Provincial Key Laboratory of Ultra High Definition Immersive Media Technology, Shenzhen Graduate School, Peking University}
  \city{Shenzhen}
\country{China}
}
\affiliation{%
\institution{Peng Cheng Laboratory}
\city{Shenzhen}
\country{China}
}

\renewcommand{\shortauthors}{Liang Xie, Yanting Li, Luyang Tang \& Wei Gao}

\begin{abstract}
  Storage and transmission challenges in dynamic 3D scene representation based on the i3DV platform, With increasing scene complexity, the explosive growth of 3D Gaussian data volume causes excessive storage space occupancy. To address this issue, we propose adopting the AVS PCRM reference software for efficient compression of Gaussian point cloud geometry data. The strategy deeply integrates the advanced encoding capabilities of AVS PCRM into the i3DV platform, forming technical complementarity with the original rate-distortion optimization mechanism based on binary hash tables. On one hand, the hash table efficiently caches inter-frame Gaussian point transformation relationships, which allows for high-fidelity transmission within a 40 Mbps bandwidth constraint. On the other hand, AVS PCRM performs precise compression on geometry data. Experimental results demonstrate that the joint framework maintains the advantages of fast rendering and high-quality synthesis in 3D Gaussian technology while achieving significant 10\%-25\% bitrate savings on universal test sets. It provides a superior rate-distortion tradeoff solution for the storage, transmission, and interaction of 3D volumetric video. 
\end{abstract}




\maketitle

\section{Introduction}

The classical 3D Gaussian method employs a large number of Gaussian points to model static scenes, with each Gaussian point containing multiple attribute values. These include position (3D coordinates), opacity (1D), covariance matrix (6D), and spherical harmonic (SH) coefficients used to represent view-dependent colors (typically using 3rd-order SH, totaling 48D). Meanwhile, a complete Gaussian point requires storing data across 59 channels~\cite{niedermayr2024compressed,bagdasarian20243dgs,huang2025hierarchical}. A moderately complex scene typically contains hundreds of thousands or even millions of Gaussian points, causing the data volume for a single frame to potentially reach hundreds of megabytes~\cite{fan2024trim}.
% (3+1+6+48+1 auxiliary parameter)

The high-dimensional, high-density point cloud data characteristic presents significant challenges for real-time compression and transmission of 3D volumetric video. On one hand, in high-resolution or complex dynamic scenes, the number of Gaussian points grows exponentially, leading to rapidly increasing storage overhead. For example, in bandwidth-constrained real-time interactive scenarios, conventional video coding schemes struggle to efficiently compress such unstructured point cloud data. Meanwhile, many experiment show that unoptimized Gaussian point cloud sequences can generate bitrate requirements exceeding 1 Gbps for 30fps volumetric video streams, which severely limits the large-scale deployment of 3D Gaussian technology in real-time immersive applications.
% \begin{table*}[h]
% \centering
% \caption{Properties of a Gauss Point}
% \begin{tabular}{|c|c|c|c|c|c|c|}
% \hline
% \textbf{Property} & \textbf{Position} & \textbf{Opaqueness} & \textbf{Size} & \textbf{Rotation} & \textbf{Spherical Harmonic Coefficient} & \textbf{Total} \\ \hline
% \textbf{Number of Channels} & 3 & 1 & 3 & 4 & 48 & 59 \\ \hline
% \end{tabular}
% \end{table*}

% To address the challenges of excessive parameters and high storage/transmission costs in volumetric video sequences, the i3DV~\cite{tang2025compressing} platform introduce an end-to-end 3D volumetric video streaming reconstruction and coding platform, which employs an anchor-based Gaussian representation method~\cite{schonberger2016structure} to construct 3D volumetric video.
% The process begins with the initial static scene at time $t_0$ from the input dynamic multi-view video using sparse point clouds obtained through Colmap~\cite{kerbl20233d, fu2024colmap} as initialization, the system performs complete scene reconstruction to generate key scenes. 
% For subsequent frames, to fully exploit temporal similarities in the scene, i3DV proposes a compact and efficient binary transform caching method to model attribute changes between anchor points across adjacent frames. This approach enables the transformation of scenes from time $t_0$ to the current timestamp, producing non-key scenes.

To address the challenges of excessive parameters and high transmission costs in volumetric video sequences, the i3DV platform~\cite{tang2025compressing,M8499} introduces an end-to-end framework for volumetric video streaming coding and reconstruction. The system employs an anchor-based Gaussian representation~\cite{schonberger2016structure} to efficiently model dynamic 3D scenes. The reconstruction process begins by initializing the static scene at time $t_0$ using sparse point clouds generated from dynamic multi-view video via Colmap~\cite{kerbl20233d,fu2024colmap}. This step produces the first set of key scenes through complete 3D reconstruction. For subsequent frames, i3DV utilizes a compact binary transform caching method to encode attribute changes between anchor points across adjacent frames, effectively exploiting temporal coherence in the scene. The approach enables efficient transformation of the reference scene at $t_0$ into non-key frames at arbitrary timestamps, significantly reducing redundancy in storage and transmission.

During frame-by-frame training, the i3DV platform jointly optimizes scene representation bitrate and reconstruction quality for optimal Rate-Distortion (RD)~\cite{guo2023rate} performance. However, the approach introduces substantial storage overhead in dynamic 3D scene modeling, especially for long sequences or HDR volumetric content. As temporal duration and scene complexity increase, the system must manage exponentially growing Gaussian point cloud data.
The storage challenge manifests in three key aspects: First, each frame requires storing millions of Gaussian points with 59-dimensional attributes. Second, dynamic modeling necessitates continuous recording of temporal information including positional offsets and attribute transformations. Third, redundant hash entries maintained for rendering quality further increase storage demands. Our tests show a mere 10-second 1080p dynamic scene can generate over 50 Gbps of raw data, severely limiting mobile and edge deployment.
To address these challenges, we present an AVS PCRM~\cite{li2024avs,wang2023rate} based compression framework that efficiently encodes geometry information. Our solution achieves a 40\% storage reduction while improving reconstruction accuracy, enabling more practical deployment of volumetric video applications.

% \vspace{-5pt}
\section{Related Work}
% Point cloud compression techniques are primarily categorized into video-based compression (V-PCC) and geometry-based compression (G-PCC), aiming to reduce data volume while preserving 3D information integrity.
Firstly, point cloud compression techniques are primarily categorized into video-based compression (V-PCC)~\cite{li2024mpeg} and geometry-based compression (G-PCC)~\cite{graziosi2020overview,li2024mpeg}, aiming to reduce data volume while preserving 3D information integrity. In addition to these mainstream approaches, the Audio Video Standard (AVS) offers specialized point cloud compression tools, such as AVS Point Cloud Reference Model (PCRM)~\cite{li2024point}, which integrates advanced intra-frame prediction, inter-frame motion modeling, and adaptive quantization to achieve competitive compression efficiency. 
% Unlike V-PCC’s reliance on 2D video codecs or G-PCC’s octree-based methods, AVS PCRM provides a hybrid solution tailored for dynamic point clouds, balancing bitrate savings with low-complexity decoding—making it particularly suitable for real-time volumetric video applications.
Meanwhile, there is also a critical need for solutions addressing both Gaussian point cloud geometry and attribute compression. Several advanced approaches have been proposed in~\cite{chen2024hac,liu2024compgs,fan2024lightgaussian,liu2024hemgs,chen2025pcgs,zhang2024gaussianimage,chen20254dgs,li2025gscodec,wang2024contextgs,liu2025compgs++,xie2022towards}.

\begin{figure*}[!t]
    \centering
    \includegraphics[width=0.95\linewidth,alt={geometry and attribute data processing pipeline}]{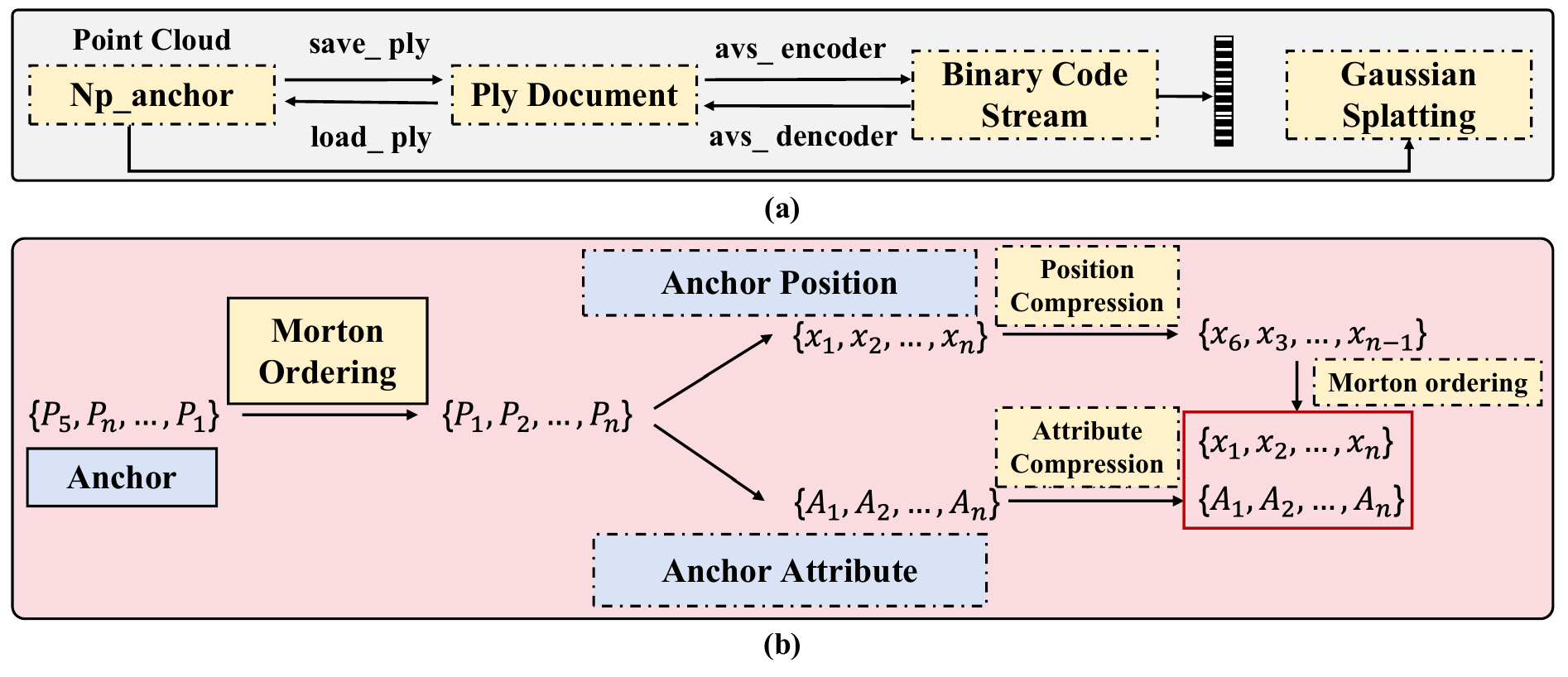}
    % \vspace{-10pt}
    \caption{The geometry and attribute data processing pipeline in i3DV~\cite{M8499} platform. (a) Data conversion and compression process. (b) Gaussian point cloud geometry and attribute data are aligned via Morton Codes~\cite{liu2024moc}.}
    \label{fig:morton}
    % \vspace{-10pt}
\end{figure*}
% \vspace{-5pt}
\subsection{Video-Based Point Cloud Compression}
In addition, V-PCC projects 3D point clouds into 2D video frames for processing via traditional video codecs, making it particularly suitable for dense, regularly distributed point clouds (\eg, RGB-D cameras).
Gao~\textit{\etal}~\cite{gao2022rate} propose a joint geometry-color distortion model that optimizes quantization parameter selection, significantly improving point cloud compression quality under target bitrate constraints.
Liu~\textit{\etal}~\cite{liu2021reduced} introduce a linear perceptual quality model for V-PCC RD optimization.
% based on geometry and color quantization step sizes. 
Their model demonstrates strong correlation with subjective quality scores while delivering superior perceptual quality at equivalent bitrates compared to conventional methods.
Li~\textit{\etal}~\cite{li2020occupancy} address compression inefficiencies caused by unoccupied pixels in V-PCC by developing an occupancy map-based RD optimization and partitioning method. 
% Their approach disregards distortion calculations for unoccupied pixels and optimizes partitioning strategies, substantially enhancing dynamic point cloud compression efficiency.
Meanwhile, Li~\textit{\etal}~\cite{li2020rate} introduce the first rate control algorithm for V-PCC by developing an optimized bit allocation framework between geometry and attribute components. Their approach eliminates bit allocation for unoccupied pixels while implementing a block-unit level parameter update mechanism through auxiliary information.
% This method achieved notable bitrate savings while improving compression performance.
Rhyu~\textit{\etal}~\cite{rhyu2020contextual} propose an enhanced V-PCC technique that preserves 3D homogeneity in 2D projections, demonstrating significant bitrate reductions in both random access and all-intra coding modes, thereby advancing point cloud compression efficacy.
V-PCC introduces edge distortion and unoccupied pixel redundancy through its 2D projection approach while exhibiting limited effectiveness for sparse point clouds compression~\cite{tu2022v}.

% \vspace{-5pt}
\subsection{Geometry-Based Point Cloud Compression}
G-PCC directly processes 3D geometry structures using methods such as octrees or predictive coding, maintaining superior geometry precision for sparse and unstructured data (\eg, LiDAR scans). For example, 
Queiroz~\textit{\etal}~\cite{de2023set} develop an embedded attribute coding method for point clouds based on the set partitioning in hierarchical trees algorithm, which integrates with the region-adaptive hierarchical transform~\cite{de2016compression} to achieve scalable compression with bitstream truncation capability. 
% Their approach demonstrates performance improvements over conventional non-predictive RAHT encoders while overcoming the coefficient prediction limitations in G-PCC standards, while preserving the critical feature of embedded coding.
Milani~\textit{\etal}~\cite{milani2020transform} propose a hybrid transform coding strategy for dynamic point clouds that combines nonlinear transforms for geometry data with linear transforms for color data, both employing region-adaptive mechanisms. By leveraging temporal redundancy for transform parameter prediction and inter-attribute prediction, their method achieves significant bitrate reduction while demonstrating superior RD performance.
Xu~\textit{\etal}~\cite{xu2020predictive} introduce a comprehensive compression framework for dynamic point cloud attributes that employs generalized graph fourier transform~\cite{sardellitti2017graph} based on gaussian markov random fields~\cite{rue2005gaussian} for optimal spatiotemporal decorrelation.
% The framework incorporates improved motion estimation methods and RD optimized coding mode selection.
Gu~\textit{\etal}~\cite{gu20203d} present a compression scheme for static voxelized point cloud attributes. Their method employs k-d tree-based geometry clustering along with an innovative graph prediction module. By leveraging representative points from encoded clusters to build predictive graph structures, the approach combines graph transforms with uniform quantization entropy coding. 
% This framework achieves enhanced RD performance while maintaining lower computational complexity.
Ramalho~\textit{\etal}~\cite{ramalho2021silhouette} enhance their previously developed silhouette-based point cloud geometry coding algorithm by introducing an innovative context selection preprocessing algorithm. 
% Their improved method tests numerous candidate context locations and selects optimal subsets for encoding.
Krivokuća~\textit{\etal}~\cite{krivokuca2021compression} propose a groundbreaking 6D representation for plenoptic point clouds, establishing a joint non-separable transform coding framework in global coordinates that unifies spatial and angular dimensions. 
% Their novel extension of RAHT and Graph Fourier Transform (GFT) to six-dimensional space accommodates both sparse/dense viewpoints and complete/incomplete plenoptic data (where incomplete data contains only visible surface point colors per viewpoint).
@inproceedings{SPCGC,
G-PCC suffers from inefficient attribute compression and suboptimal temporal redundancy exploitation in dynamic scenarios, with geometry quantization often causing surface artifacts~\cite{wei2025high,gao2025deep,PKU-DPCC,xie2024pchmvision,xie2024learningpcc,xie2024roi,SAVD-PCGC,PDNet,xie2022end,xie2025learning,xie2022online,gao2022openpointcloud,wu2024adaptive}.

% 在压缩高斯点云序列过程中，采用原始i3DV平台编码与i3DV加入AVS PCRM压缩第一帧（i帧）数据几何坐标的压缩性能、客观质量、编解码器时间比较
\begin{table*}[!t]
\centering
\renewcommand\arraystretch{1.3} 
\setlength{\tabcolsep}{0.26mm}
\caption{Comparison of compression performance, objective quality, and encoding/decoding time between original i3DV~\cite{tang2025compressing} and AVS PCRM~\cite{li2024avs} enhanced i3DV for Gaussian point cloud sequences.}
% \vspace{-10pt}
\begin{tabular}{lccccccccc}
\toprule
\textbf{Methods} & \textbf{PSNR} & \textbf{SSIM} & \textbf{LPIPS} & \textbf{Size (MB)} & \textbf{Total Rate} & \textbf{XYZ Size (MB)} & \textbf{XYZ Rate} & \textbf{Enc Time (s)} & \textbf{Dec Time (s)} \\
\midrule\midrule

& \multicolumn{9}{c}{Dance\_Dunhuang\_Pair\_1080} \\
\cmidrule(lr){2-10}
i3DV2.0~\cite{tang2025compressing} & 39.52 & 0.980 & 0.077 & 0.8324 & 0.00\% & 0.1431 & 0.00\% & 1.5832 & 1.5896 \\
i3DV2.0 + AVS PCRM~\cite{li2024avs} & 39.52 & 0.980 & 0.077 & 0.748 & 10.14\% & 0.0588 & 58.91\% & 1.7082 & 1.4963 \\
\midrule
 & \multicolumn{9}{c}{Show\_Groups\_4K}\\
 \cmidrule(lr){2-10}
i3DV2.0~\cite{tang2025compressing} & 37.27 & 0.944 & 0.230 & 0.9171 & 0.00\% & 0.2209 & 0.00\% & 1.4107 & 2.1839 \\
i3DV2.0 + AVS PCRM~\cite{li2024avs} & 37.27 & 0.944 & 0.230 & 0.7942 & 13.40\% & 0.0979 & 55.68\% & 1.5068 & 2.5599 \\
\midrule
 & \multicolumn{9}{c}{VRU\_dg4}\\
 \cmidrule(lr){2-10}
i3DV2.0~\cite{tang2025compressing} & 29.10 & 0.924 & 0.167 & 3.3453 & 0.00\% & 1.2519 & 0.00\% & 6.663 & 8.8349 \\
i3DV2.0 + AVS PCRM~\cite{li2024avs} & 29.10 & 0.924 & 0.167 & 2.4876 & 25.64\% & 0.3928 & 68.62\% & 11.3854 & 9.6462 \\
\midrule
 & \multicolumn{9}{c}{VRU\_gz}\\
 \cmidrule(lr){2-10}
i3DV2.0~\cite{tang2025compressing} & 31.63 & 0.949 & 0.197 & 2.9555 & 0.00\% & 1.1578 & 0.00\% & 8.0172 & 6.414 \\
i3DV2.0 + AVS PCRM~\cite{li2024avs} & 31.63 & 0.949 & 0.197 & 2.2394 & 24.23\% & 0.4407 & 61.94\% & 9.1977 & 8.6551 \\
\bottomrule
\end{tabular}
\label{tab:objective_metrics}
% \vspace{-10pt}
\end{table*}

% \vspace{-5pt}
\section{Methodology}
\subsection{Data Conversion and Lossless Compression}

This paper presents a comprehensive point cloud data processing pipeline to address the data format compatibility issues between the i3DV platform and the AVS PCRM reference software. As shown in Fig.~\ref{fig:morton} (a), our solution implements a three-stage workflow: First, the anchor data in numpy array format from the i3DV platform is converted into standard PLY format, with the "save\_ply" module accurately extracting both geometry features and attribute information from the point cloud. The core encoding phase then employs the "avs\_encoder" module to efficiently compress the PLY formatted point cloud data. Following attribute data encoding in the i3DV platform, the "avs\_decoder" module precisely reconstructs the geometry data. In the final reconstruction stage, the system converts the decoded data back to numpy format by "load\_ply", performs rigorous geometry-attribute alignment, and ultimately achieves high-quality image reconstruction through Gaussian splatting. The entire process, with clearly marked data flow directions, not only ensures complete data conversion integrity but also strictly maintains the precision requirements of the original anchor data, providing a reliable geometry foundation for subsequent rendering tasks. 
% Experimental results demonstrate that this solution significantly improves encoding efficiency while preserving data accuracy, offering an effective approach to data interoperability challenges in heterogeneous systems.

% \vspace{-5pt}
\subsection{Geometry and Attributes Alignment}

During the point cloud compression process, when the AVS PCRM software employs lossless compression for geometry data, an inconsistency arises between the positional relationships of point clouds before and after encoding. The phenomenon leads to mismatches between the decoded geometry structure and the original data, consequently affecting the correct correspondence of Gaussian point cloud geometry and attribute data (\eg, color information).
To address this issue, we propose a spatial sorting optimization scheme based on Morton Code. Specifically, prior to encoding, we first perform Morton Code sorting on the geometry data to establish stable spatial indexing relationships. After encoding is completed, the decoder reconstructs the geometry data according to the same Morton order, ensuring the point cloud positions perfectly match the original data. The alignment process is shown in Fig.~\ref{fig:morton} (b), all related point cloud attribute data are rearranged according to the Morton order index values, thereby guaranteeing strict correspondence between Gaussian geometry and attributes data. the solution not only effectively resolves the positional inconsistency before and after encoding by introducing a spatial sorting mechanism, but also significantly improves data processing reliability and decoding accuracy, providing an accurate data foundation for subsequent point cloud rendering and analysis.

% \vspace{-5pt}
\section{Experiment}
The paper implements the AVS PCRM (v11)\footnote{\url{https://git.openi.org.cn/xiel/AVS-PCRM}} point cloud geometry compression tool on the i3DV (v2.0) platform\footnote{\url{https://github.com/Pomelomm/iFVC}} as a replacement for the original scheme, conducting comprehensive subjective and objective performance evaluation experiments across four standard datasets. The following presents the comparative results of both objective metrics and subjective quality assessments.

% \vspace{-5pt}
\subsection{Test sequence}
% The test sequences include a total of 4 scenes, each containing 250 frames of data at a frame rate of 25 fps. Among these, 2 scenes are from the multi-view video dataset VRU-Basketball based on basketball matches, featuring many foreground characters with fast movements, large scenes, sparse camera arrangements, including 34 camera views covering an angle close to 90 degrees, with a resolution of 1920×1080. The other 2 scenes are from the multi-view 3D human reconstruction dataset PKU-MVHumans~\cite{zheng2024pku}, covering single or multiple high-definition dynamic characters, showcasing a rich variety of human body shapes, movements, and clothing details, where one scene includes 360-degree panoramic views captured by 60 cameras, with a resolution of 1920×1080, and the other scene consists of 56 cameras forming a 270-degree view, with a resolution of 3840×2160.
% Based on the data provided, we conduct a systematic evaluation of four point cloud datasets (Dance\_Dunhuang\_Pair\_1080, Show\_Groups\_4K, VRU\_dg4, and VRU\_gz) under different methods (i3Dv2.0 original encoding and i3DV2.0+AVS\_PCRM). 

Particularly, our experiment evaluation utilize four distinct test sequences, each containing 250 frames at 25 fps. The test dataset comprises two complementary components representing different capture scenarios. First, we include two challenging basketball sequences from the VRU-Basketball~\cite{VRU2024} multi-view video dataset, featuring fast-moving players in large-scale environments captured by 34 cameras arranged in a sparse 90-degree configuration at 1920×1080 resolution. Second, we incorporate two high-fidelity human reconstruction sequences from the PKU-MVHumans dataset~\cite{zheng2024pku}, including one 360-degree panoramic scene (60 cameras at 1920×1080) and one 270-degree scene (56 cameras at 3840×2160), which showcase diverse human subjects with detailed motion and clothing textures.
For comprehensive evaluation, we conduct systematic comparisons across four representative point cloud datasets: Dance\_Dunhuang\_Pair\_1080, Show\_Groups\_4K, VRU\_dg4, and VRU\_gz. Each dataset is processed using two distinct encoding approaches, the baseline i3DV2.0 original encoding framework and our proposed i3DV2.0 plus AVS PCRM enhanced compression method to enable rigorous performance benchmarking.

\begin{figure}[!t]
    \centering
    \includegraphics[width=1\linewidth]{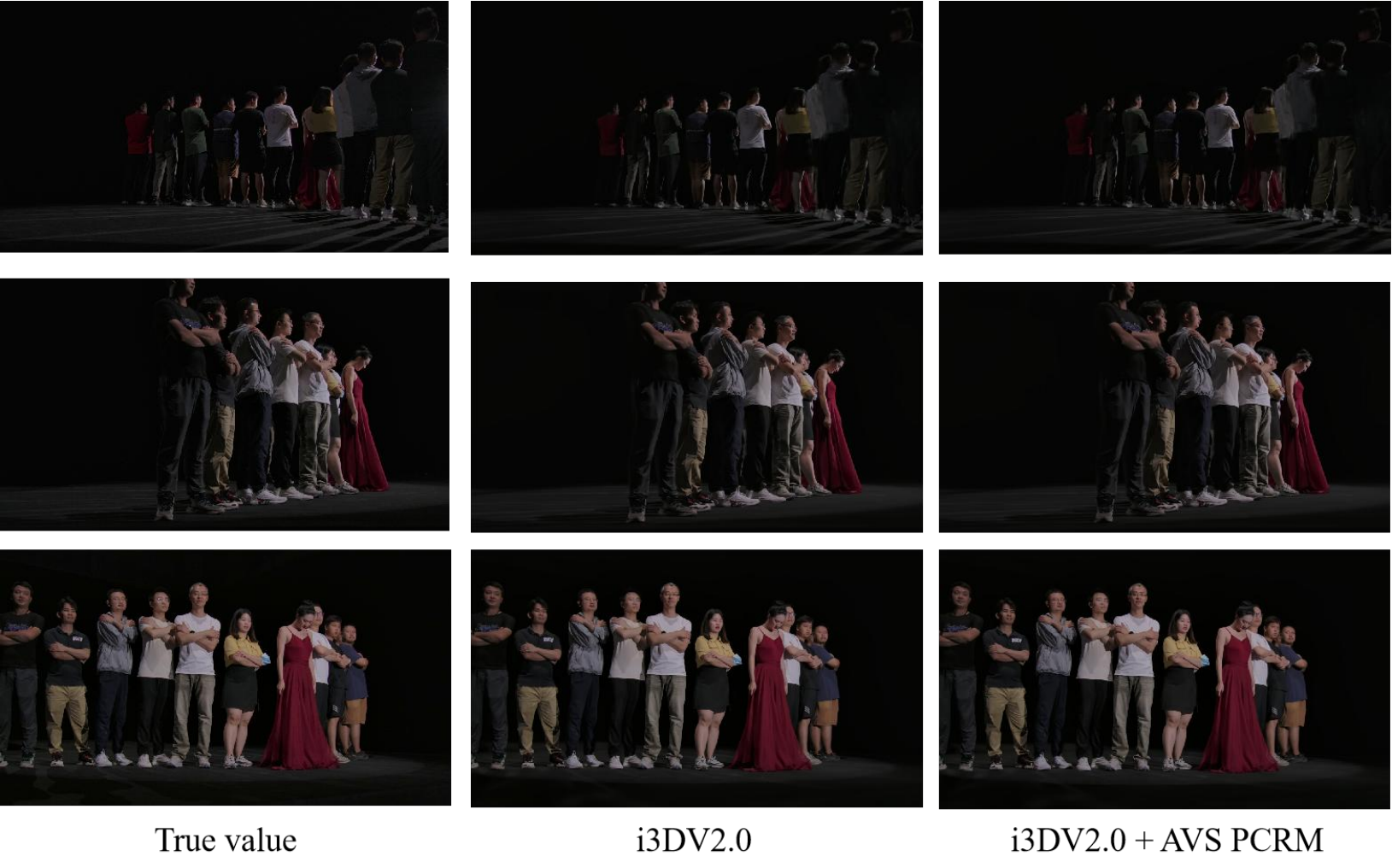}
    % \vspace{-15pt}
    \caption{The reconstruction results of different algorithms under the Show\_Groups\_4K dataset.}
    \label{fig:show2}
    % \vspace{-10pt}
    % \vspace{-17pt}
\end{figure}

% \vspace{-5pt}
\subsection{Performance Comparison}
% in Table~\ref{tab:objective_metrics}

\textbf{PSNR Analysis:} In the Dance\_Dunhuang\_Pair\_1080 dataset, the PSNR of i3DV2.0 software is 39.52, and for AVS PCRM it is 39.52, showing comparable performance. In Show\_Groups\\\_4K, i3DV2.0 scores 37.27, and AVS PCRM also scores 37.27, maintaining a tie. In the VRU dataset, the PSNR for i3DV2.0 in versions VRU\_dg4 and VRU\_gz are 29.10 and 31.63 respectively, while for AVS PCRM they are 29.10 and 31.63 respectively, indicating consistent performance in PSNR between the two methods with no significant difference observed.

\textbf{SSIM Analysis:} The SSIM values~\cite{zhao2015ssim} reflect structural similarity. For Dance\_Dunhuang\_Pair\_1080, both i3DV2.0 and AVS PCRM have SSIM values of 0.980, in Show\_Groups\\\_4K both have 0.944, and for VRU\_dg4 and VRU\_gz they are 0.924 and 0.949 respectively, demonstrating stable performance in structural fidelity between the two methods with minimal difference.

\textbf{LPIPS Analysis:} The LPIPS~\cite{zhang2018unreasonable} measures perceptual similarity. For Dance\_Dunhuang\_Pair\_1080, both i3DV2.0 and AVS PCRM have LPIPS values of 0.077, in Show\_Groups\_4K both have 0.230, and for VRU\_dg4 and VRU\_gz they are 0.167 and 0.197 respectively, indicating a small difference in perceptual quality between the two methods, with a slight improvement in the VRU\_gz version.

\textbf{Compression Rate:} The Size for Dance\_Dunhuang\_Pair\\\_1080 is 0.834 MB for both methods with a compression rate of 10.14\%. For Show\_Groups\_4K it is 0.971 MB with a compression rate~\cite{yang2023lossy} of 13.40\%. For VRU\_dg4 and VRU\_gz they are 3.343 MB (25.64\%) and 2.394 MB (24.23\%) respectively. AVS PCRM has a smaller file size on VRU g\_gz, showing slightly better compression efficiency.

\textbf{Encoding and Decoding Time:} Meanwhile, the encoding and decoding time show little difference between the two methods. For Dance\_Dunhuang\_Pair\_1080, the encoding time is 1.582s and decoding time is 1.589s for both. For Show\_Groups\_4K, the encoding time is 1.410s with a slight difference in decoding time (2.183s vs 2.559s). For VRU\_dg4, the encoding time is 6.663s and decoding time is 8.834s, while for VRU\_gz they are 8.017s and 6.414s respectively, showing a slight advantage for AVS PCRM in decoding efficiency.

\begin{figure}[!t]
    \centering
    \includegraphics[width=1\linewidth]{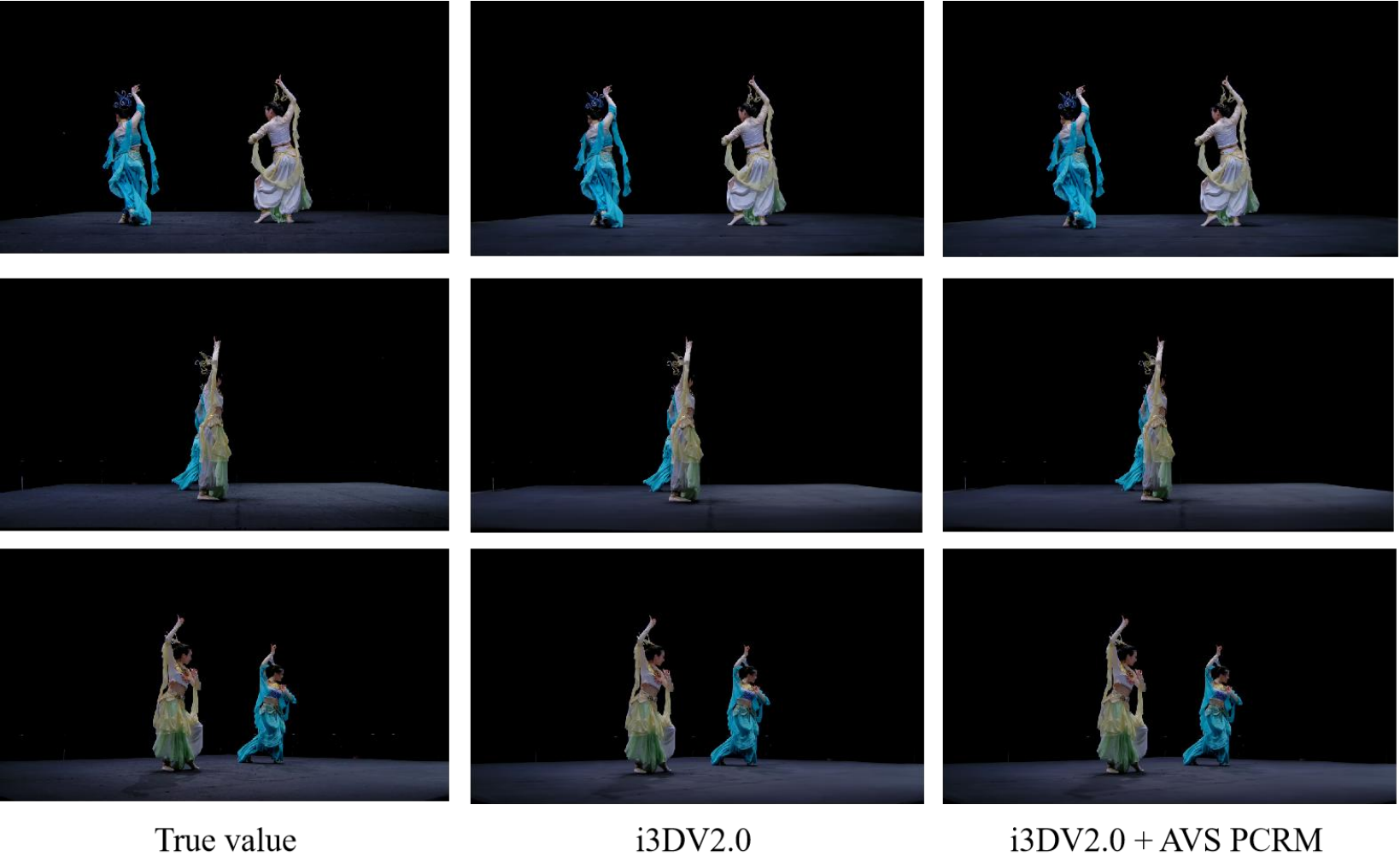}
    \caption{The reconstruction results of different algorithms under the Dance\_Dunhuang\_Pair dataset.}
    \label{fig:show1}
\end{figure}

% \vspace{-5pt}
\subsection{Datasets Differences Comparison}

\textbf{Dance\_Dunhuang\_Pair\_1080 Dataset:} The PSNR and SSIM values are relatively high, indicating good quality of this dataset. Both methods show consistent compression effects, with XYZ saving a bit rate of 58.91\%, and encoding and decoding times are relatively short (1.5-1.6 seconds), making it suitable for efficient processing.

\textbf{Show\_Groups\_4K Dataset:} The PSNR and SSIM values are slightly lower, with a larger file size (0.971 MB) and a compression rate of 13.40\%. The XYZ bit rate saving is 55.68\%, and the decoding time is slightly longer (approximately 2.2-2.6 seconds), reflecting an increase in processing complexity due to high-resolution data.
    
\textbf{VRU Dataset:} For VRU\_dg4 and VRU\_gz, the PSNR values are relatively low (29.10 and 31.63 respectively), with file sizes of 3.343 MB and 2.394 MB respectively. The compression rates are higher (25.64\% and 24.23\%), and the XYZ bit rate savings reach 68.62\% and 61.94\%. The encoding and decoding times are longer (6-8 seconds), suggesting that this dataset may contain more complex geometry structures, requiring higher compression and decoding demands.

% \textbf{Method comparison}
In conclusion, the two methods show almost identical performance in PSNR, SSIM, and LPIPS, indicating no significant difference in quality assessment metrics. AVS PCRM has a slight advantage in file size and decoding time for the VRU\_gz version, likely due to its optimized decoding algorithm. In terms of encoding time, there is no significant difference between the two methods, making them suitable for real-time processing needs across different datasets. AVS PCRM has a slight edge in XYZ bit rate savings (\eg, 58.91\% for Dance\_Dunhuang\_Pair\_1080 vs 0.00\%), suggesting it may be more efficient in spatial compression. Meanwhile, Fig.~\ref{fig:show2}, Fig.~\ref{fig:show3}, Fig.~\ref{fig:show1} and Fig.~\ref{fig:show4} present a subjective comparison experiment between the i3DV platform and the AVS PCRM method as an alternative point cloud compression scheme. Visual inspection reveals that there is no significant difference in rendering quality between the compressed model and the original model.

\begin{figure}[!t]
    \centering
    \includegraphics[width=1\linewidth,alt={pic5}]{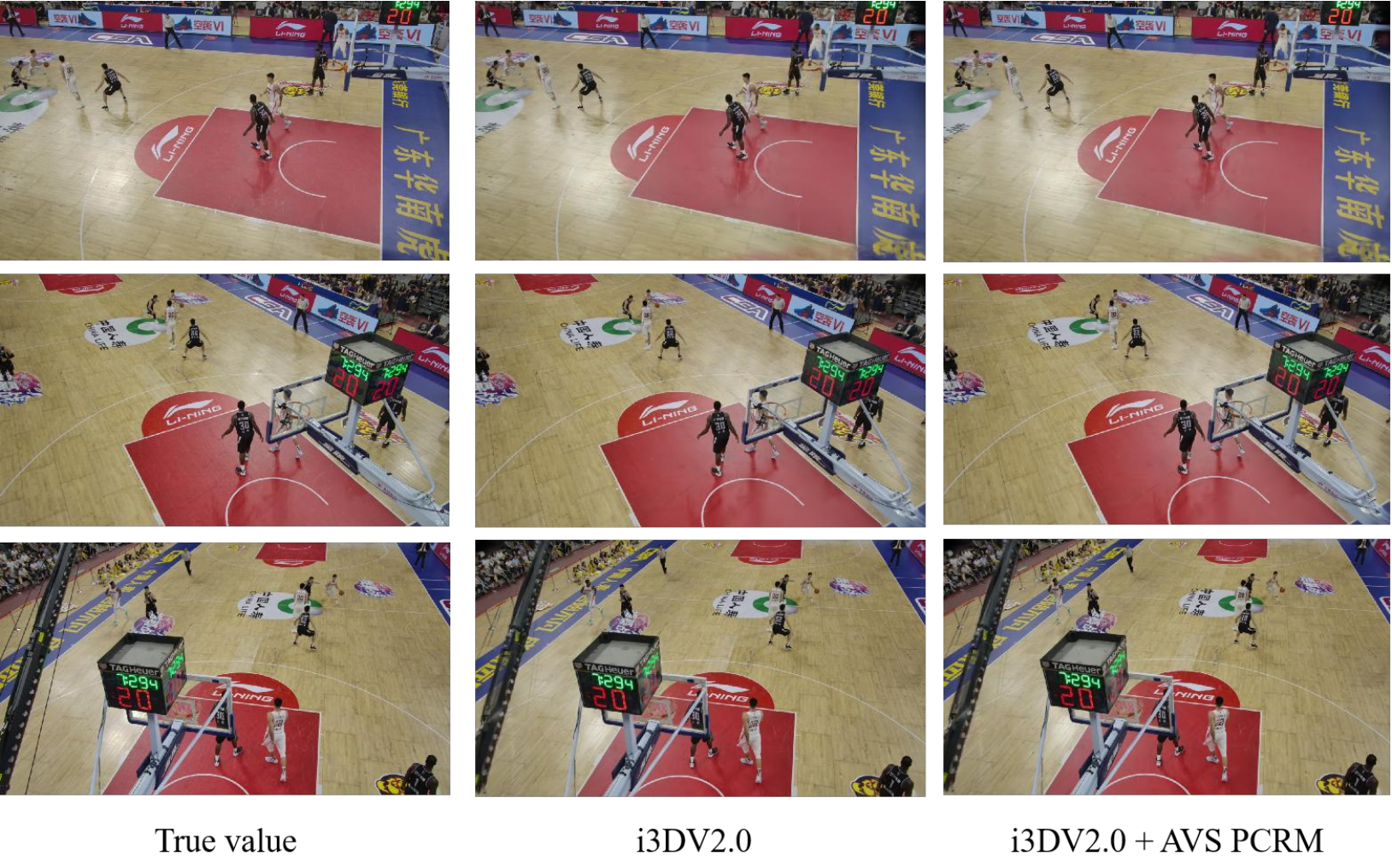}
    % \vspace{-15pt}
    \caption{The reconstruction results of different algorithms under the VRU\_dg4 dataset.}
    \label{fig:show3}
    % \vspace{-10pt}
\end{figure}

\begin{figure}[!t]
    \centering
    \includegraphics[width=1\linewidth]{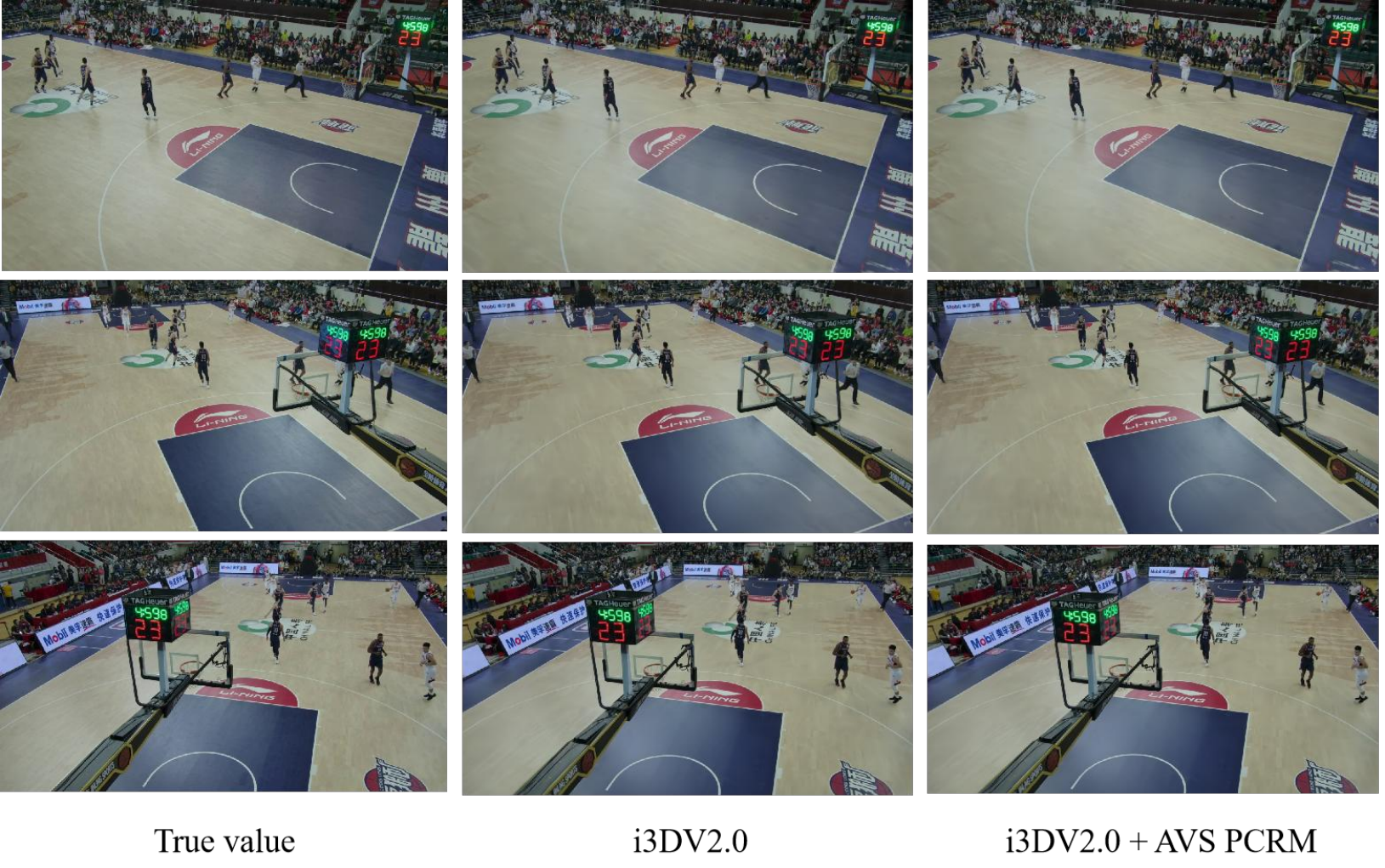}
    \caption{The reconstruction results of different algorithms under the VRU\_gz dataset.}
    \label{fig:show4}
\end{figure}

% \vspace{-5pt}
\section{Conclusion}
% The paper innovatively introduces the adoption of AVS PCRM reference software for efficient compression of Gaussian point cloud geometry data. By deeply integrating the advanced encoding capabilities of AVS PCRM into the i3DV platform, we establish technical complementarity with the existing binary hash table-based RD optimization mechanism:
% (1) The hash table efficiently caches inter-frame Gaussian point transformation relationships (enabling high-fidelity transmission at 40Mbps bandwidth)
% (2) Simultaneously, AVS PCRM performs precise compression on geometric data
% % Experimental results demonstrate that the unified framework maintains the fast rendering and high-quality synthesis advantages of 3D Gaussian technology while achieving significant bitrate savings of 10\%-25\% on standard test sets. This approach provides an improved rate-distortion trade-off solution for 3D volumetric video storage, transmission, and interactive applications.

The paper innovatively introduces the adoption of AVS PCRM reference software for efficient compression of Gaussian point cloud geometry data. By deeply integrating the advanced encoding capabilities of AVS PCRM into the i3DV platform, we establish technical complementarity with the existing binary hash table-based RD optimization mechanism. First, the hash table efficiently caches inter-frame Gaussian point transformation relationships, enabling high-fidelity transmission at 40 Mbps bandwidth. Second, AVS PCRM performs precise compression on geometry data, ensuring further bitrate efficiency.
Experimental results demonstrate that the unified framework maintains the fast rendering and high-quality synthesis advantages of 3D Gaussian technology while achieving significant bitrate savings of 10\%–25\% on standard test sets. 
% The approach provides an improved rate-distortion trade-off solution for 3D volumetric video storage, transmission, and interactive applications.

% \section{Acknowledgements}

\bibliographystyle{ACM-Reference-Format}

\bibliography{sample-base}


\begin{thebibliography}{60}

%%% the Web address.

\ifx \showCODEN    \undefined \def \showCODEN     #1{\unskip}     \fi
\ifx \showISBNx    \undefined \def \showISBNx     #1{\unskip}     \fi
\ifx \showISBNxiii \undefined \def \showISBNxiii  #1{\unskip}     \fi
\ifx \showISSN     \undefined \def \showISSN      #1{\unskip}     \fi
\ifx \showLCCN     \undefined \def \showLCCN      #1{\unskip}     \fi
\ifx \shownote     \undefined \def \shownote      #1{#1}          \fi
\ifx \showarticletitle \undefined \def \showarticletitle #1{#1}   \fi
\ifx \showURL      \undefined \def \showURL       {\relax}        \fi
% invisible to TeX
\providecommand\bibfield[2]{#2}
\providecommand\bibinfo[2]{#2}
\providecommand\natexlab[1]{#1}
\providecommand\showeprint[2][]{arXiv:#2}

\bibitem[Bagdasarian et~al\mbox{.}(2024)]%
        {bagdasarian20243dgs}
\bibfield{author}{\bibinfo{person}{Milena~T Bagdasarian}, \bibinfo{person}{Paul Knoll}, \bibinfo{person}{Y Li}, \bibinfo{person}{Florian Barthel}, \bibinfo{person}{Anna Hilsmann}, \bibinfo{person}{Peter Eisert}, {and} \bibinfo{person}{Wieland Morgenstern}.} \bibinfo{year}{2024}\natexlab{}.
\newblock \showarticletitle{3dgs.zip: A survey on 3d gaussian splatting compression methods}. In \bibinfo{booktitle}{\emph{Computer Graphics Forum}}. Wiley Online Library, \bibinfo{pages}{e70078}.
\newblock


\bibitem[Chen et~al\mbox{.}(2025b)]%
        {chen2025pcgs}
\bibfield{author}{\bibinfo{person}{Yihang Chen}, \bibinfo{person}{Mengyao Li}, \bibinfo{person}{Qianyi Wu}, \bibinfo{person}{Weiyao Lin}, \bibinfo{person}{Mehrtash Harandi}, {and} \bibinfo{person}{Jianfei Cai}.} \bibinfo{year}{2025}\natexlab{b}.
\newblock \showarticletitle{Pcgs: Progressive compression of 3d gaussian splatting}.
\newblock \bibinfo{journal}{\emph{arXiv preprint arXiv:2503.08511}} (\bibinfo{year}{2025}).
\newblock


\bibitem[Chen et~al\mbox{.}(2024)]%
        {chen2024hac}
\bibfield{author}{\bibinfo{person}{Yihang Chen}, \bibinfo{person}{Qianyi Wu}, \bibinfo{person}{Weiyao Lin}, \bibinfo{person}{Mehrtash Harandi}, {and} \bibinfo{person}{Jianfei Cai}.} \bibinfo{year}{2024}\natexlab{}.
\newblock \showarticletitle{Hac: Hash-grid assisted context for 3d gaussian splatting compression}. In \bibinfo{booktitle}{\emph{European Conference on Computer Vision}}. Springer, \bibinfo{pages}{422--438}.
\newblock


\bibitem[Chen et~al\mbox{.}(2025a)]%
        {chen20254dgs}
\bibfield{author}{\bibinfo{person}{Zicong Chen}, \bibinfo{person}{Zhenghao Chen}, \bibinfo{person}{Wei Jiang}, \bibinfo{person}{Wei Wang}, \bibinfo{person}{Lei Liu}, {and} \bibinfo{person}{Dong Xu}.} \bibinfo{year}{2025}\natexlab{a}.
\newblock \showarticletitle{4DGS-CC: A Contextual Coding Framework for 4D Gaussian Splatting Data Compression}.
\newblock \bibinfo{journal}{\emph{arXiv preprint arXiv:2504.18925}} (\bibinfo{year}{2025}).
\newblock


\bibitem[de~Queiroz et~al\mbox{.}(2023)]%
        {de2023set}
\bibfield{author}{\bibinfo{person}{Ricardo de Queiroz}, \bibinfo{person}{Andre Souto}, \bibinfo{person}{Victor Figueiredo}, {and} \bibinfo{person}{Philip Chou}.} \bibinfo{year}{2023}\natexlab{}.
\newblock \showarticletitle{Set Partitioning in Hierarchical Trees for Point Cloud Attribute Compression}.
\newblock \bibinfo{journal}{\emph{Authorea Preprints}} (\bibinfo{year}{2023}).
\newblock


\bibitem[De~Queiroz and Chou(2016)]%
        {de2016compression}
\bibfield{author}{\bibinfo{person}{Ricardo~L De~Queiroz} {and} \bibinfo{person}{Philip~A Chou}.} \bibinfo{year}{2016}\natexlab{}.
\newblock \showarticletitle{Compression of 3D point clouds using a region-adaptive hierarchical transform}.
\newblock \bibinfo{journal}{\emph{IEEE Transactions on Image Processing}} \bibinfo{volume}{25}, \bibinfo{number}{8} (\bibinfo{year}{2016}), \bibinfo{pages}{3947--3956}.
\newblock


\bibitem[Fan et~al\mbox{.}(2024b)]%
        {fan2024trim}
\bibfield{author}{\bibinfo{person}{Lue Fan}, \bibinfo{person}{Yuxue Yang}, \bibinfo{person}{Minxing Li}, \bibinfo{person}{Hongsheng Li}, {and} \bibinfo{person}{Zhaoxiang Zhang}.} \bibinfo{year}{2024}\natexlab{b}.
\newblock \showarticletitle{Trim 3d gaussian splatting for accurate geometry representation}.
\newblock \bibinfo{journal}{\emph{arXiv preprint arXiv:2406.07499}} (\bibinfo{year}{2024}).
\newblock


\bibitem[Fan et~al\mbox{.}(2024a)]%
        {fan2024lightgaussian}
\bibfield{author}{\bibinfo{person}{Zhiwen Fan}, \bibinfo{person}{Kevin Wang}, \bibinfo{person}{Kairun Wen}, \bibinfo{person}{Zehao Zhu}, \bibinfo{person}{Dejia Xu}, \bibinfo{person}{Zhangyang Wang}, {et~al\mbox{.}}} \bibinfo{year}{2024}\natexlab{a}.
\newblock \showarticletitle{Lightgaussian: Unbounded 3d gaussian compression with 15x reduction and 200+ fps}.
\newblock \bibinfo{journal}{\emph{Advances in neural information processing systems}}  \bibinfo{volume}{37} (\bibinfo{year}{2024}), \bibinfo{pages}{140138--140158}.
\newblock


\bibitem[Fu et~al\mbox{.}(2024)]%
        {fu2024colmap}
\bibfield{author}{\bibinfo{person}{Yang Fu}, \bibinfo{person}{Sifei Liu}, \bibinfo{person}{Amey Kulkarni}, \bibinfo{person}{Jan Kautz}, \bibinfo{person}{Alexei~A Efros}, {and} \bibinfo{person}{Xiaolong Wang}.} \bibinfo{year}{2024}\natexlab{}.
\newblock \showarticletitle{Colmap-free 3d gaussian splatting}. In \bibinfo{booktitle}{\emph{IEEE/CVF Conference on Computer Vision and Pattern Recognition}}. \bibinfo{pages}{20796--20805}.
\newblock


\bibitem[Gao et~al\mbox{.}(2022a)]%
        {gao2022rate}
\bibfield{author}{\bibinfo{person}{Pan Gao}, \bibinfo{person}{Shengzhou Luo}, {and} \bibinfo{person}{Manoranjan Paul}.} \bibinfo{year}{2022}\natexlab{a}.
\newblock \showarticletitle{Rate-distortion modeling for bit rate constrained point cloud compression}.
\newblock \bibinfo{journal}{\emph{IEEE Transactions on Circuits and Systems for Video Technology}} \bibinfo{volume}{33}, \bibinfo{number}{5} (\bibinfo{year}{2022}), \bibinfo{pages}{2424--2438}.
\newblock


\bibitem[Gao et~al\mbox{.}(2025)]%
        {gao2025deep}
\bibfield{author}{\bibinfo{person}{Wei Gao}, \bibinfo{person}{Liang Xie}, \bibinfo{person}{Songlin Fan}, {and} \bibinfo{person}{Ge Li}.} \bibinfo{year}{2025}\natexlab{}.
\newblock \showarticletitle{Deep Learning-Based Point Cloud Compression: An In-Depth Survey and Benchmark}.
\newblock \bibinfo{journal}{\emph{IEEE Transactions on Pattern Analysis and Machine Intelligence}} (\bibinfo{year}{2025}).
\newblock


\bibitem[Gao et~al\mbox{.}(2022b)]%
        {gao2022openpointcloud}
\bibfield{author}{\bibinfo{person}{Wei Gao}, \bibinfo{person}{Hua Ye}, \bibinfo{person}{Ge Li}, \bibinfo{person}{Huiming Zheng}, \bibinfo{person}{Yuyang Wu}, {and} \bibinfo{person}{Liang Xie}.} \bibinfo{year}{2022}\natexlab{b}.
\newblock \showarticletitle{{OpenPointCloud: An Open-Source Algorithm Library of Deep Learning Based Point Cloud Compression}}. In \bibinfo{booktitle}{\emph{ACM International Conference on Multimedia}}. \bibinfo{pages}{7347--7350}.
\newblock


\bibitem[Graziosi et~al\mbox{.}(2020)]%
        {graziosi2020overview}
\bibfield{author}{\bibinfo{person}{Danillo Graziosi}, \bibinfo{person}{Ohji Nakagami}, \bibinfo{person}{Satoru Kuma}, \bibinfo{person}{Alexandre Zaghetto}, \bibinfo{person}{Teruhiko Suzuki}, {and} \bibinfo{person}{Ali Tabatabai}.} \bibinfo{year}{2020}\natexlab{}.
\newblock \showarticletitle{An overview of ongoing point cloud compression standardization activities: Video-based (V-PCC) and geometry-based (G-PCC)}.
\newblock \bibinfo{journal}{\emph{APSIPA Transactions on Signal and Information Processing}}  \bibinfo{volume}{9} (\bibinfo{year}{2020}), \bibinfo{pages}{e13}.
\newblock


\bibitem[Gu et~al\mbox{.}(2020)]%
        {gu20203d}
\bibfield{author}{\bibinfo{person}{Shuai Gu}, \bibinfo{person}{Junhui Hou}, \bibinfo{person}{Huanqiang Zeng}, {and} \bibinfo{person}{Hui Yuan}.} \bibinfo{year}{2020}\natexlab{}.
\newblock \showarticletitle{3D point cloud attribute compression via graph prediction}.
\newblock \bibinfo{journal}{\emph{IEEE Signal Processing Letters}}  \bibinfo{volume}{27} (\bibinfo{year}{2020}), \bibinfo{pages}{176--180}.
\newblock


\bibitem[Guo et~al\mbox{.}(2023)]%
        {guo2023rate}
\bibfield{author}{\bibinfo{person}{Tian Guo}, \bibinfo{person}{Hui Yuan}, \bibinfo{person}{Lu Wang}, {and} \bibinfo{person}{Tingting Wang}.} \bibinfo{year}{2023}\natexlab{}.
\newblock \showarticletitle{Rate-distortion optimized quantization for geometry-based point cloud compression}.
\newblock \bibinfo{journal}{\emph{Journal of Electronic Imaging}} \bibinfo{volume}{32}, \bibinfo{number}{1} (\bibinfo{year}{2023}), \bibinfo{pages}{013047--013047}.
\newblock


\bibitem[Huang et~al\mbox{.}(2025)]%
        {huang2025hierarchical}
\bibfield{author}{\bibinfo{person}{He Huang}, \bibinfo{person}{Wenjie Huang}, \bibinfo{person}{Qi Yang}, \bibinfo{person}{Yiling Xu}, {and} \bibinfo{person}{Zhu Li}.} \bibinfo{year}{2025}\natexlab{}.
\newblock \showarticletitle{A hierarchical compression technique for 3d gaussian splatting compression}. In \bibinfo{booktitle}{\emph{IEEE International Conference on Acoustics, Speech and Signal Processing}}. IEEE, \bibinfo{pages}{1--5}.
\newblock


\bibitem[Kerbl et~al\mbox{.}(2023)]%
        {kerbl20233d}
\bibfield{author}{\bibinfo{person}{Bernhard Kerbl}, \bibinfo{person}{Georgios Kopanas}, \bibinfo{person}{Thomas Leimk{\"u}hler}, {and} \bibinfo{person}{George Drettakis}.} \bibinfo{year}{2023}\natexlab{}.
\newblock \showarticletitle{3d gaussian splatting for real-time radiance field rendering}.
\newblock \bibinfo{journal}{\emph{ACM Trans. Graph.}} \bibinfo{volume}{42}, \bibinfo{number}{4} (\bibinfo{year}{2023}), \bibinfo{pages}{139--1}.
\newblock


\bibitem[Krivoku{\'c}a et~al\mbox{.}(2021)]%
        {krivokuca2021compression}
\bibfield{author}{\bibinfo{person}{Maja Krivoku{\'c}a}, \bibinfo{person}{Ehsan Miandji}, \bibinfo{person}{Christine Guillemot}, {and} \bibinfo{person}{Philip~A Chou}.} \bibinfo{year}{2021}\natexlab{}.
\newblock \showarticletitle{Compression of plenoptic point cloud attributes using 6-D point clouds and 6-D transforms}.
\newblock \bibinfo{journal}{\emph{IEEE Transactions on Multimedia}}  \bibinfo{volume}{25} (\bibinfo{year}{2021}), \bibinfo{pages}{593--607}.
\newblock


\bibitem[Li et~al\mbox{.}(2024a)]%
        {li2024avs}
\bibfield{author}{\bibinfo{person}{Ge Li}, \bibinfo{person}{Wei Gao}, {and} \bibinfo{person}{Wen Gao}.} \bibinfo{year}{2024}\natexlab{a}.
\newblock \showarticletitle{AVS point cloud compression standard}.
\newblock In \bibinfo{booktitle}{\emph{Point Cloud Compression: Technologies and Standardization}}. \bibinfo{publisher}{Springer}, \bibinfo{pages}{167--197}.
\newblock


\bibitem[Li et~al\mbox{.}(2024b)]%
        {li2024mpeg}
\bibfield{author}{\bibinfo{person}{Ge Li}, \bibinfo{person}{Wei Gao}, {and} \bibinfo{person}{Wen Gao}.} \bibinfo{year}{2024}\natexlab{b}.
\newblock \showarticletitle{MPEG geometry-based point cloud compression (G-PCC) standard}.
\newblock In \bibinfo{booktitle}{\emph{Point Cloud Compression: Technologies and Standardization}}. \bibinfo{publisher}{Springer}, \bibinfo{pages}{135--165}.
\newblock


\bibitem[Li et~al\mbox{.}(2024c)]%
        {li2024point}
\bibfield{author}{\bibinfo{person}{Ge Li}, \bibinfo{person}{Wei Gao}, {and} \bibinfo{person}{Wen Gao}.} \bibinfo{year}{2024}\natexlab{c}.
\newblock \bibinfo{booktitle}{\emph{Point Cloud Compression: Technologies and Standardization}}.
\newblock \bibinfo{publisher}{Springer Nature}.
\newblock


\bibitem[Li et~al\mbox{.}(2020a)]%
        {li2020occupancy}
\bibfield{author}{\bibinfo{person}{Li Li}, \bibinfo{person}{Zhu Li}, {and} \bibinfo{person}{Shan Liu}.} \bibinfo{year}{2020}\natexlab{a}.
\newblock \showarticletitle{Occupancy-map-based rate distortion optimization and partition for video-based point cloud compression}.
\newblock \bibinfo{journal}{\emph{IEEE Transactions on Circuits and Systems for Video Technology}} \bibinfo{volume}{31}, \bibinfo{number}{1} (\bibinfo{year}{2020}), \bibinfo{pages}{326--338}.
\newblock


\bibitem[Li et~al\mbox{.}(2020b)]%
        {li2020rate}
\bibfield{author}{\bibinfo{person}{Li Li}, \bibinfo{person}{Zhu Li}, \bibinfo{person}{Shan Liu}, {and} \bibinfo{person}{Houqiang Li}.} \bibinfo{year}{2020}\natexlab{b}.
\newblock \showarticletitle{Rate control for video-based point cloud compression}.
\newblock \bibinfo{journal}{\emph{IEEE Transactions on Image Processing}}  \bibinfo{volume}{29} (\bibinfo{year}{2020}), \bibinfo{pages}{6237--6250}.
\newblock


\bibitem[Li et~al\mbox{.}(2025)]%
        {li2025gscodec}
\bibfield{author}{\bibinfo{person}{Sicheng Li}, \bibinfo{person}{Chengzhen Wu}, \bibinfo{person}{Hao Li}, \bibinfo{person}{Xiang Gao}, \bibinfo{person}{Yiyi Liao}, {and} \bibinfo{person}{Lu Yu}.} \bibinfo{year}{2025}\natexlab{}.
\newblock \showarticletitle{GSCodec Studio: A Modular Framework for Gaussian Splat Compression}.
\newblock \bibinfo{journal}{\emph{arXiv preprint arXiv:2506.01822}} (\bibinfo{year}{2025}).
\newblock


\bibitem[Liu et~al\mbox{.}(2024a)]%
        {liu2024hemgs}
\bibfield{author}{\bibinfo{person}{Lei Liu}, \bibinfo{person}{Zhenghao Chen}, \bibinfo{person}{Wei Jiang}, \bibinfo{person}{Wei Wang}, {and} \bibinfo{person}{Dong Xu}.} \bibinfo{year}{2024}\natexlab{a}.
\newblock \showarticletitle{Hemgs: A hybrid entropy model for 3d gaussian splatting data compression}.
\newblock \bibinfo{journal}{\emph{arXiv preprint arXiv:2411.18473}} (\bibinfo{year}{2024}).
\newblock


\bibitem[Liu et~al\mbox{.}(2021)]%
        {liu2021reduced}
\bibfield{author}{\bibinfo{person}{Qi Liu}, \bibinfo{person}{Hui Yuan}, \bibinfo{person}{Raouf Hamzaoui}, \bibinfo{person}{Honglei Su}, \bibinfo{person}{Junhui Hou}, {and} \bibinfo{person}{Huan Yang}.} \bibinfo{year}{2021}\natexlab{}.
\newblock \showarticletitle{Reduced reference perceptual quality model with application to rate control for video-based point cloud compression}.
\newblock \bibinfo{journal}{\emph{IEEE Transactions on Image Processing}}  \bibinfo{volume}{30} (\bibinfo{year}{2021}), \bibinfo{pages}{6623--6636}.
\newblock


\bibitem[Liu et~al\mbox{.}(2024b)]%
        {liu2024moc}
\bibfield{author}{\bibinfo{person}{Xueyuan Liu}, \bibinfo{person}{Zhuoran Song}, \bibinfo{person}{Hao Chen}, {and} \bibinfo{person}{Xing Li}.} \bibinfo{year}{2024}\natexlab{b}.
\newblock \showarticletitle{MoC: A Morton-Code-Based Fine-Grained Quantization for Accelerating Point Cloud Neural Networks}. In \bibinfo{booktitle}{\emph{ACM/IEEE Design Automation Conference}}. \bibinfo{pages}{1--6}.
\newblock


\bibitem[Liu et~al\mbox{.}(2025)]%
        {liu2025compgs++}
\bibfield{author}{\bibinfo{person}{Xiangrui Liu}, \bibinfo{person}{Xinju Wu}, \bibinfo{person}{Shiqi Wang}, \bibinfo{person}{Zhu Li}, {and} \bibinfo{person}{Sam Kwong}.} \bibinfo{year}{2025}\natexlab{}.
\newblock \showarticletitle{CompGS++: Compressed Gaussian Splatting for Static and Dynamic Scene Representation}.
\newblock \bibinfo{journal}{\emph{arXiv preprint arXiv:2504.13022}} (\bibinfo{year}{2025}).
\newblock


\bibitem[Liu et~al\mbox{.}(2024c)]%
        {liu2024compgs}
\bibfield{author}{\bibinfo{person}{Xiangrui Liu}, \bibinfo{person}{Xinju Wu}, \bibinfo{person}{Pingping Zhang}, \bibinfo{person}{Shiqi Wang}, \bibinfo{person}{Zhu Li}, {and} \bibinfo{person}{Sam Kwong}.} \bibinfo{year}{2024}\natexlab{c}.
\newblock \showarticletitle{Compgs: Efficient 3d scene representation via compressed gaussian splatting}. In \bibinfo{booktitle}{\emph{ACM International Conference on Multimedia}}. \bibinfo{pages}{2936--2944}.
\newblock


\bibitem[Milani et~al\mbox{.}(2020)]%
        {milani2020transform}
\bibfield{author}{\bibinfo{person}{Simone Milani}, \bibinfo{person}{Enrico Polo}, {and} \bibinfo{person}{Simone Limuti}.} \bibinfo{year}{2020}\natexlab{}.
\newblock \showarticletitle{A transform coding strategy for dynamic point clouds}.
\newblock \bibinfo{journal}{\emph{IEEE Transactions on Image Processing}}  \bibinfo{volume}{29} (\bibinfo{year}{2020}), \bibinfo{pages}{8213--8225}.
\newblock


\bibitem[Niedermayr et~al\mbox{.}(2024)]%
        {niedermayr2024compressed}
\bibfield{author}{\bibinfo{person}{Simon Niedermayr}, \bibinfo{person}{Josef Stumpfegger}, {and} \bibinfo{person}{R{\"u}diger Westermann}.} \bibinfo{year}{2024}\natexlab{}.
\newblock \showarticletitle{Compressed 3d gaussian splatting for accelerated novel view synthesis}. In \bibinfo{booktitle}{\emph{IEEE/CVF Conference on Computer Vision and Pattern Recognition}}. \bibinfo{pages}{10349--10358}.
\newblock


\bibitem[Ramalho et~al\mbox{.}(2021)]%
        {ramalho2021silhouette}
\bibfield{author}{\bibinfo{person}{Evaristo Ramalho}, \bibinfo{person}{Eduardo Peixoto}, {and} \bibinfo{person}{Edil Medeiros}.} \bibinfo{year}{2021}\natexlab{}.
\newblock \showarticletitle{Silhouette 4D with context selection: Lossless geometry compression of dynamic point clouds}.
\newblock \bibinfo{journal}{\emph{IEEE Signal Processing Letters}}  \bibinfo{volume}{28} (\bibinfo{year}{2021}), \bibinfo{pages}{1660--1664}.
\newblock


\bibitem[Rhyu et~al\mbox{.}(2020)]%
        {rhyu2020contextual}
\bibfield{author}{\bibinfo{person}{Sungryeul Rhyu}, \bibinfo{person}{Junsik Kim}, \bibinfo{person}{Jiheon Im}, {and} \bibinfo{person}{Kyuheon Kim}.} \bibinfo{year}{2020}\natexlab{}.
\newblock \showarticletitle{Contextual homogeneity-based patch decomposition method for higher point cloud compression}.
\newblock \bibinfo{journal}{\emph{IEEE Access}}  \bibinfo{volume}{8} (\bibinfo{year}{2020}), \bibinfo{pages}{207805--207812}.
\newblock


\bibitem[Rue and Held(2005)]%
        {rue2005gaussian}
\bibfield{author}{\bibinfo{person}{Havard Rue} {and} \bibinfo{person}{Leonhard Held}.} \bibinfo{year}{2005}\natexlab{}.
\newblock \bibinfo{booktitle}{\emph{Gaussian Markov random fields: theory and applications}}.
\newblock \bibinfo{publisher}{Chapman and Hall/CRC}.
\newblock


\bibitem[Sardellitti et~al\mbox{.}(2017)]%
        {sardellitti2017graph}
\bibfield{author}{\bibinfo{person}{Stefania Sardellitti}, \bibinfo{person}{Sergio Barbarossa}, {and} \bibinfo{person}{Paolo Di~Lorenzo}.} \bibinfo{year}{2017}\natexlab{}.
\newblock \showarticletitle{On the graph Fourier transform for directed graphs}.
\newblock \bibinfo{journal}{\emph{IEEE Journal of Selected Topics in Signal Processing}} \bibinfo{volume}{11}, \bibinfo{number}{6} (\bibinfo{year}{2017}), \bibinfo{pages}{796--811}.
\newblock


\bibitem[Schonberger and Frahm(2016)]%
        {schonberger2016structure}
\bibfield{author}{\bibinfo{person}{Johannes~L Schonberger} {and} \bibinfo{person}{Jan-Michael Frahm}.} \bibinfo{year}{2016}\natexlab{}.
\newblock \showarticletitle{Structure-from-motion revisited}. In \bibinfo{booktitle}{\emph{IEEE conference on computer vision and pattern recognition}}. \bibinfo{pages}{4104--4113}.
\newblock


\bibitem[Tang et~al\mbox{.}(2025)]%
        {tang2025compressing}
\bibfield{author}{\bibinfo{person}{Luyang Tang}, \bibinfo{person}{Jiayu Yang}, \bibinfo{person}{Rui Peng}, {and} \bibinfo{person}{Yongqi Zhai}.} \bibinfo{year}{2025}\natexlab{}.
\newblock \showarticletitle{Compressing streamable free-viewpoint videos to 0.1 mb per frame}. In \bibinfo{booktitle}{\emph{AAAI Conference on Artificial Intelligence}}, Vol.~\bibinfo{volume}{39}. \bibinfo{pages}{7257--7265}.
\newblock


\bibitem[Tang et~al\mbox{.}(2024)]%
        {M8499}
\bibfield{author}{\bibinfo{person}{Luyang Tang}, \bibinfo{person}{Jiayu Yang}, \bibinfo{person}{Rui Peng}, {and} \bibinfo{person}{Yongqi Zhai}.} \bibinfo{year}{August, 2024}\natexlab{}.
\newblock \showarticletitle{i3DV: Intelligent 3D Volumetric Video Encoding Platform}.
\newblock \bibinfo{journal}{\emph{AVS, M8499, Online}} (\bibinfo{year}{August, 2024}).
\newblock


\bibitem[Tu et~al\mbox{.}(2022)]%
        {tu2022v}
\bibfield{author}{\bibinfo{person}{Renwei Tu}, \bibinfo{person}{Gangyi Jiang}, \bibinfo{person}{Mei Yu}, \bibinfo{person}{Ting Luo}, \bibinfo{person}{Zongju Peng}, {and} \bibinfo{person}{Fen Chen}.} \bibinfo{year}{2022}\natexlab{}.
\newblock \showarticletitle{V-PCC projection based blind point cloud quality assessment for compression distortion}.
\newblock \bibinfo{journal}{\emph{IEEE Transactions on Emerging Topics in Computational Intelligence}} \bibinfo{volume}{7}, \bibinfo{number}{2} (\bibinfo{year}{2022}), \bibinfo{pages}{462--473}.
\newblock


\bibitem[VRU(2024)]%
        {VRU2024}
\bibfield{author}{\bibinfo{person}{VRU}.} \bibinfo{year}{2024}\natexlab{}.
\newblock \bibinfo{title}{VRU-sequence}.
\newblock
\urldef\tempurl%
\url{https://anonymous.4open.science/r/VRU-Sequence/}
\showURL{%
\tempurl}


\bibitem[Wang et~al\mbox{.}(2023)]%
        {wang2023rate}
\bibfield{author}{\bibinfo{person}{Yang Wang}, \bibinfo{person}{Wei Gao}, {and} \bibinfo{person}{Xingming Mu}.} \bibinfo{year}{2023}\natexlab{}.
\newblock \showarticletitle{Rate control optimization for joint geometry and attribute coding of lidar point clouds}. In \bibinfo{booktitle}{\emph{IEEE International Conference on Visual Communications and Image Processing}}. IEEE, \bibinfo{pages}{1--5}.
\newblock


\bibitem[Wang et~al\mbox{.}(2024)]%
        {wang2024contextgs}
\bibfield{author}{\bibinfo{person}{Yufei Wang}, \bibinfo{person}{Zhihao Li}, \bibinfo{person}{Lanqing Guo}, \bibinfo{person}{Wenhan Yang}, \bibinfo{person}{Alex Kot}, {and} \bibinfo{person}{Bihan Wen}.} \bibinfo{year}{2024}\natexlab{}.
\newblock \showarticletitle{Contextgs: Compact 3d gaussian splatting with anchor level context model}.
\newblock \bibinfo{journal}{\emph{Advances in neural information processing systems}}  \bibinfo{volume}{37} (\bibinfo{year}{2024}), \bibinfo{pages}{51532--51551}.
\newblock


\bibitem[Wei et~al\mbox{.}(2025)]%
        {wei2025high}
\bibfield{author}{\bibinfo{person}{Yuxuan Wei}, \bibinfo{person}{Zehan Wang}, \bibinfo{person}{Tian Guo}, \bibinfo{person}{Hao Liu}, \bibinfo{person}{Liquan Shen}, {and} \bibinfo{person}{Hui Yuan}.} \bibinfo{year}{2025}\natexlab{}.
\newblock \showarticletitle{High efficiency Wiener filter-based point cloud quality enhancement for MPEG G-PCC}.
\newblock \bibinfo{journal}{\emph{IEEE Transactions on Circuits and Systems for Video Technology}} (\bibinfo{year}{2025}).
\newblock


\bibitem[Wu et~al\mbox{.}(2024)]%
        {wu2024adaptive}
\bibfield{author}{\bibinfo{person}{Yuyang Wu}, \bibinfo{person}{Liang Xie}, \bibinfo{person}{Shangkun Sun}, \bibinfo{person}{Wei Gao}, {and} \bibinfo{person}{Yiqiang Yan}.} \bibinfo{year}{2024}\natexlab{}.
\newblock \showarticletitle{Adaptive Intra Period Size for Deep Learning-based Screen Content Video Coding}. In \bibinfo{booktitle}{\emph{IEEE International Conference on Multimedia and Expo Workshops}}. IEEE, \bibinfo{pages}{1--6}.
\newblock


\bibitem[Xie and Fan(2025)]%
        {xie2025learning}
\bibfield{author}{\bibinfo{person}{Liang Xie} {and} \bibinfo{person}{Songlin Fan}.} \bibinfo{year}{2025}\natexlab{}.
\newblock \showarticletitle{A Learning-based Multi-Frame Visual Feature Framework for Real-Time Driver Fatigue Detection}. In \bibinfo{booktitle}{\emph{Conference of the Nations of the Americas Chapter of the Association for Computational Linguistics: Human Language Technologies}}. \bibinfo{pages}{61--69}.
\newblock


\bibitem[Xie and Gao(2024a)]%
        {xie2024learningpcc}
\bibfield{author}{\bibinfo{person}{Liang Xie} {and} \bibinfo{person}{Wei Gao}.} \bibinfo{year}{2024}\natexlab{a}.
\newblock \showarticletitle{LearningPCC: A PyTorch Library for Learning-Based Point Cloud Compression}. In \bibinfo{booktitle}{\emph{ACM International Conference on Multimedia}}. \bibinfo{pages}{11234--11238}.
\newblock


\bibitem[Xie and Gao(2024b)]%
        {xie2024pchmvision}
\bibfield{author}{\bibinfo{person}{Liang Xie} {and} \bibinfo{person}{Wei Gao}.} \bibinfo{year}{2024}\natexlab{b}.
\newblock \showarticletitle{PCHMVision: An Open-Source Library of Point Cloud Compression for Human and Machine Vision}. In \bibinfo{booktitle}{\emph{ACM International Conference on Multimedia}}. \bibinfo{pages}{11239--11243}.
\newblock


\bibitem[Xie et~al\mbox{.}(2024a)]%
        {PDNet}
\bibfield{author}{\bibinfo{person}{Liang Xie}, \bibinfo{person}{Wei Gao}, \bibinfo{person}{Songlin Fan}, {and} \bibinfo{person}{Zhaojian Yao}.} \bibinfo{year}{2024}\natexlab{a}.
\newblock \showarticletitle{PDNet: Parallel Dual-branch Network for Point Cloud Geometry Compression and Analysis}. In \bibinfo{booktitle}{\emph{Data Compression Conference}}. \bibinfo{pages}{596--596}.
\newblock


\bibitem[Xie et~al\mbox{.}(2022a)]%
        {xie2022end}
\bibfield{author}{\bibinfo{person}{Liang Xie}, \bibinfo{person}{Wei Gao}, {and} \bibinfo{person}{Huiming Zheng}.} \bibinfo{year}{2022}\natexlab{a}.
\newblock \showarticletitle{{End-to-End Point Cloud Geometry Compression and Analysis with Sparse Tensor}}. In \bibinfo{booktitle}{\emph{International Workshop on Advances in Point Cloud Compression, Processing and Analysis}}. \bibinfo{pages}{27--32}.
\newblock


\bibitem[Xie et~al\mbox{.}(2024b)]%
        {xie2024roi}
\bibfield{author}{\bibinfo{person}{Liang Xie}, \bibinfo{person}{Wei Gao}, \bibinfo{person}{Huiming Zheng}, {and} \bibinfo{person}{Ge Li}.} \bibinfo{year}{2024}\natexlab{b}.
\newblock \showarticletitle{ROI-Guided Point Cloud Geometry Compression Towards Human and Machine Vision}. In \bibinfo{booktitle}{\emph{ACM International Conference on Multimedia}}. \bibinfo{pages}{3741--3750}.
\newblock


\bibitem[Xie et~al\mbox{.}(2024c)]%
        {SAVD-PCGC}
\bibfield{author}{\bibinfo{person}{Liang Xie}, \bibinfo{person}{Wei Gao}, \bibinfo{person}{Huiming Zheng}, {and} \bibinfo{person}{Hua Ye}.} \bibinfo{year}{2024}\natexlab{c}.
\newblock \showarticletitle{Semantic-Aware Visual Decomposition for Point Cloud Geometry Compression}. In \bibinfo{booktitle}{\emph{Data Compression Conference}}. \bibinfo{pages}{595--595}.
\newblock


\bibitem[Xie et~al\mbox{.}(2022b)]%
        {xie2022online}
\bibfield{author}{\bibinfo{person}{Liang Xie}, \bibinfo{person}{MengHao Hu}, {and} \bibinfo{person}{XinBei Bai}.} \bibinfo{year}{2022}\natexlab{b}.
\newblock \showarticletitle{{Online Improved Vehicle Tracking Accuracy via Unsupervised Route Generation}}. In \bibinfo{booktitle}{\emph{IEEE 34th International Conference on Tools with Artificial Intelligence}}. IEEE, \bibinfo{pages}{788--792}.
\newblock


\bibitem[Xie et~al\mbox{.}(2022c)]%
        {xie2022towards}
\bibfield{author}{\bibinfo{person}{Liang Xie}, \bibinfo{person}{MengHao Hu}, {and} \bibinfo{person}{XinBei Bai}.} \bibinfo{year}{2022}\natexlab{c}.
\newblock \showarticletitle{{Towards Hardware-Friendly and Robust Facial Landmark Detection Method}}. In \bibinfo{booktitle}{\emph{International Conference on Neural Information Processing}}. Springer, \bibinfo{pages}{432--444}.
\newblock


\bibitem[Xie et~al\mbox{.}(2024d)]%
        {PKU-DPCC}
\bibfield{author}{\bibinfo{person}{Liang Xie}, \bibinfo{person}{Xingming Mu}, \bibinfo{person}{Ge Li}, \bibinfo{person}{Wei Gao}, {et~al\mbox{.}}} \bibinfo{year}{2024}\natexlab{d}.
\newblock \showarticletitle{PKU-DPCC: A New Dataset for Dynamic Point Cloud Compression}.
\newblock \bibinfo{journal}{\emph{APSIPA Transactions on Signal and Information Processing}} \bibinfo{volume}{13}, \bibinfo{number}{6} (\bibinfo{year}{2024}).
\newblock


\bibitem[Xu et~al\mbox{.}(2020)]%
        {xu2020predictive}
\bibfield{author}{\bibinfo{person}{Yiqun Xu}, \bibinfo{person}{Wei Hu}, \bibinfo{person}{Shanshe Wang}, \bibinfo{person}{Xinfeng Zhang}, \bibinfo{person}{Shiqi Wang}, \bibinfo{person}{Siwei Ma}, \bibinfo{person}{Zongming Guo}, {and} \bibinfo{person}{Wen Gao}.} \bibinfo{year}{2020}\natexlab{}.
\newblock \showarticletitle{Predictive generalized graph Fourier transform for attribute compression of dynamic point clouds}.
\newblock \bibinfo{journal}{\emph{IEEE Transactions on Circuits and Systems for Video Technology}} \bibinfo{volume}{31}, \bibinfo{number}{5} (\bibinfo{year}{2020}), \bibinfo{pages}{1968--1982}.
\newblock


\bibitem[Yang and Mandt(2023)]%
        {yang2023lossy}
\bibfield{author}{\bibinfo{person}{Ruihan Yang} {and} \bibinfo{person}{Stephan Mandt}.} \bibinfo{year}{2023}\natexlab{}.
\newblock \showarticletitle{Lossy image compression with conditional diffusion models}.
\newblock \bibinfo{journal}{\emph{Advances in Neural Information Processing Systems}}  \bibinfo{volume}{36} (\bibinfo{year}{2023}), \bibinfo{pages}{64971--64995}.
\newblock


\bibitem[Zhang et~al\mbox{.}(2018)]%
        {zhang2018unreasonable}
\bibfield{author}{\bibinfo{person}{Richard Zhang}, \bibinfo{person}{Phillip Isola}, \bibinfo{person}{Alexei~A Efros}, \bibinfo{person}{Eli Shechtman}, {and} \bibinfo{person}{Oliver Wang}.} \bibinfo{year}{2018}\natexlab{}.
\newblock \showarticletitle{The unreasonable effectiveness of deep features as a perceptual metric}. In \bibinfo{booktitle}{\emph{IEEE conference on computer vision and pattern recognition}}. \bibinfo{pages}{586--595}.
\newblock


\bibitem[Zhang et~al\mbox{.}(2024)]%
        {zhang2024gaussianimage}
\bibfield{author}{\bibinfo{person}{Xinjie Zhang}, \bibinfo{person}{Xingtong Ge}, \bibinfo{person}{Tongda Xu}, \bibinfo{person}{Dailan He}, \bibinfo{person}{Yan Wang}, \bibinfo{person}{Hongwei Qin}, \bibinfo{person}{Guo Lu}, \bibinfo{person}{Jing Geng}, {and} \bibinfo{person}{Jun Zhang}.} \bibinfo{year}{2024}\natexlab{}.
\newblock \showarticletitle{Gaussianimage: 1000 fps image representation and compression by 2d gaussian splatting}. In \bibinfo{booktitle}{\emph{European Conference on Computer Vision}}. Springer, \bibinfo{pages}{327--345}.
\newblock


\bibitem[Zhao et~al\mbox{.}(2015)]%
        {zhao2015ssim}
\bibfield{author}{\bibinfo{person}{Tiesong Zhao}, \bibinfo{person}{Jiheng Wang}, \bibinfo{person}{Zhou Wang}, {and} \bibinfo{person}{Chang~Wen Chen}.} \bibinfo{year}{2015}\natexlab{}.
\newblock \showarticletitle{SSIM-based coarse-grain scalable video coding}.
\newblock \bibinfo{journal}{\emph{IEEE Transactions on Broadcasting}} \bibinfo{volume}{61}, \bibinfo{number}{2} (\bibinfo{year}{2015}), \bibinfo{pages}{210--221}.
\newblock


\bibitem[Zheng et~al\mbox{.}(2024)]%
        {zheng2024pku}
\bibfield{author}{\bibinfo{person}{Xiaoyun Zheng}, \bibinfo{person}{Liwei Liao}, \bibinfo{person}{Xufeng Li}, \bibinfo{person}{Jianbo Jiao}, \bibinfo{person}{Rongjie Wang}, \bibinfo{person}{Feng Gao}, \bibinfo{person}{Shiqi Wang}, {and} \bibinfo{person}{Ronggang Wang}.} \bibinfo{year}{2024}\natexlab{}.
\newblock \showarticletitle{Pku-dymvhumans: A multi-view video benchmark for high-fidelity dynamic human modeling}. In \bibinfo{booktitle}{\emph{IEEE/CVF Conference on Computer Vision and Pattern Recognition}}. \bibinfo{pages}{22530--22540}.
\newblock


\end{thebibliography}
\end{document}